\documentclass[aps,prl,groupedaddress]{revtex4}
\usepackage{amsfonts,amssymb}
\usepackage{amsthm}
\usepackage{tikz}
\usetikzlibrary{calc}
\usetikzlibrary{arrows}
\usepackage{tikz-3dplot}
\usepackage{color} 
\usepackage[fleqn]{amsmath}
\usepackage{amsthm}
\usepackage{amssymb}
\usepackage{amsfonts}
\usepackage{bbm}
\usepackage{epsfig}
\usepackage{times}
\usepackage{hyperref}
\usepackage{bm}
\usepackage{float} 
\usepackage{appendix} 
\usepackage{mathrsfs}
\usepackage{appendix}

\usepackage{graphicx}

\DeclareGraphicsExtensions{.png}

\newcommand{\comment}[1]{}

\usepackage{natbib}
\theoremstyle{plain}

\theoremstyle{definition}

\begin{document}

\title{Are There Testable Discrete Poincar\'e Invariant Physical Theories?}

\author{Adrian \surname{Kent}} \affiliation{Centre for
	Quantum Information and Foundations, DAMTP, Centre for Mathematical
	Sciences, University of Cambridge, Wilberforce Road, Cambridge, CB3
	0WA, U.K.}  \affiliation{Perimeter Institute for Theoretical
	Physics, 31 Caroline Street North, Waterloo, ON N2L 2Y5, Canada.}
	\date{\today}

\begin{abstract} 
In a model of physics taking place on a discrete set of points that approximates Minkowski space,
one might perhaps expect there to be an empirically identifiable preferred frame. 
However, the work of Dowker, Bombelli, Henson, and
Sorkin\cite{dhs,bhs} might be taken to suggest that random sprinklings of points in
Minkowski space define a discrete model that is provably
Poincar\'e invariant in a natural sense. 
We examine this possibility
here.   

We argue that a genuinely Poincar\'e invariant model requires a 
probability distribution on sprinklable sets -- Poincar\'e orbits of
sprinklings -- rather than individual sprinklings. 
The corresponding $\sigma$-algebra contains only sets of measure zero
or one.   This makes testing the hypothesis of discrete Poincar\'e 
invariance problematic, since any local violation of Poincar\'e 
invariance, however gross and large scale, is possible, and cannot
be said to be improbable.   
 
We also note that the Bombelli-Henson-Sorkin \cite{bhs} argument,
which rules out constructions of preferred timelike directions 
for typical sprinklings, is not sufficient to establish full Lorentz
invariance.  For example, once a pair of timelike separated points 
is fixed, a preferred spacelike direction {\it can} be defined for a typical sprinkling,
breaking the remaining rotational invariance. 

\end{abstract}
\maketitle
	
\section{Introduction: Dowker, Bombelli, Henson and Sorkin on sprinklings}

It seems very plausible that modelling space-time by a real manifold
is an idealization.   Indeed, much work on quantum gravity
starts by hypothesizing that a fundamental theory of
space-time physics involves discrete structures, whether spin
networks, causal sets, or other mathematical objects.   

It could be that this view is correct but only
becomes interesting and coherent in the context of a theory that 
includes gravity, in which the manifold that approximately describes
the large-scale structure of space-time turns out to be better 
described on small scales by a discrete point set with appropriate
mathematical relations defined between pairs (or among subsets) of
points.   But one might also expect or hope that the theory makes sense without
gravity, giving us a discrete version of special
relativistic physics in Minkowski space.   
One reason for hoping this is that it is natural to look at discrete
approximations to Minkowski space in order to understand whether discreteness is compatible with Lorentz and
Poincar\'e invariance or whether these are necessarily broken.   
Defining the question precisely already raises subtle issues in
this context; it seems much harder to give a meaningful definition
for general Lorentzian manifolds. 
Another is that ``discrete relativistic field theory in
Minkowski space'' might be much simpler to define than, and also a good
stepping stone towards, ``discrete quantum gravity''.    
Ideas in this direction have recently been presented by Bedingham
\cite{bedingham2018collapse}.
Our discussion also applies to discrete approximations to
more cosmologically relevant manifolds with continuous global
symmetries.  

Hossenfelder has shown that there are no Poincar\'e invariant networks
with locally finite distributions of nodes and links in Minkowski
space \cite{0264-9381-32-20-207001}.   This leaves open the
possibility that discrete theories might be defined by 
point sets approximating Minkowski space, either by requiring no
network structure or by allowing nodes to have infinitely many 
links.\footnote{For example, we can define a network from a 
sprinkling by joining every pair of timelike separated points.
In such a network, every point almost surely has infinitely many links.}

So, are there Poincar\'e or Lorentz invariant
discrete point set structures that approximate Minkowski space?   
For the Lorentz group, this is a special case of the general question -- are there Lorentz
invariant discrete structures that approximate Lorentzian manifolds? --
considered by Dowker, Henson
and Sorkin \cite{dhs} (DHS) in a pioneering paper that sets out the foundations of causal set theory.
As we understand it, the existence or otherwise of discrete
approximations to Minkowski space is ultimately not very important for
the causal set programme, which seeks rules for generating discrete
structures that are approximately consistent with the phenomenology
of general relativity and Big Bang cosmology.     
Our discussion here, which focusses on discretizations of Minkowski
space, is indebted to work on causal sets, 
but not in any sense a critique of the main thrust of that programme. 

As DHS note, there is a sense in which the answer to both questions is clearly negative, 
since no discrete structure can be invariant under the action 
of every element of the continuous Lorentz group: 
\begin{quote}
Naturally, there can be no question of a literal action of the entire Lorentz group on an individual
discrete structure.\cite{dhs}
\end{quote}

However, they argue that this is not the physically relevant sense: 
\begin{quote}
Rather such a structure can only be Lorentz invariant in the same sense that a
fluid is translation invariant. This should not detract from the fact that a fluid is indeed translation
invariant in an important sense, whereas a crystalline solid is not.\cite{dhs}
\end{quote}

They go on to offer a physical criterion: 
\begin{quote}
What does it mean to say that a discrete theory respects Lorentz invariance? It
is difficult to give a precise answer, but intuitively the import is clear. Whenever a
continuum is a good approximation to the underlying structure (and
assuming specifically 
that the approximating continuum is a Lorentzian manifold M), the underlying
discreteness must not, in and of itself, suffice to distinguish a local Lorentz frame at
any point of M. In consequence, no phenomenological theory in M derived from such
a scheme can involve a local (or global) Lorentz frame either. \cite{dhs}
\end{quote}

This motivates the defininition of a {\it sprinkling} (here taken to have Planck density) as: 	
\begin{quote}
a Poisson process. To see what this
means, imagine dividing $M$, using any local coordinate systems, into small boxes of
volume $V$, and then placing a ``sprinkled point'' independently into each box with
probability $V /V_{\rm fund}$, where $V_{\rm  fund}$ is the fundamental volume (of order the Planck
volume). The Poisson process is the limit of this procedure as $V$ tends to zero.
Because spacetime volume is an invariant, the limiting process is independent of the
coordinate systems used to define the boxes. It follows that one cannot tell which
frame was used to produce the sprinkling: the approximation is
``equally good in all
frames''.  \cite{dhs} 
\end{quote}

DHS make the following claim: 
\begin{quote}
We want to emphasise that not only is the process of sprinkling Lorentz invariant
but so also are almost all of the individual causets that are
generated. \cite{dhs}
\end{quote}
As far as I understand it, their discussion of this point appeals to the
intuition that the Lorentz invariance of the sprinkling process
makes it seem plausible that almost all individual sprinklings are 
Lorentz invariant, together with the observation 
that some common objections (based, for example,
on the existence of voids in sprinklings) can be refuted.  

\subsection{The BHS theorem}

Bombelli, Henson and Sorkin \cite{bhs} (BHS) take the discussion
further: 
\begin{quote}
[ Ref. \cite{dhs} ] presented strong evidence that causets produced by
sprinkling into Minkowski spacetime meet this criterion, but a skeptic could still have
found grounds for doubt. In this paper, we prove a theorem that we believe removes most
of the remaining doubt. \cite{bhs}
\end{quote}

BHS underline that a Lorentz invariant definition of a probability
distribution on point sets does {\em not}, per se, logically imply Lorentz invariance of a 
typical set: 
\begin{quotation}
The fact that the process of ``causet sprinkling'' in Minkowski space is Lorentz invariant
is an important first step in the argument. (In this process we include both the Poisson
sprinkling as such and the subsequent induction of the causal order. Both steps are
manifestly Lorentz invariant since they depend only on the volume element and the causal
structure of the spacetime, respectively). But Lorentz invariance of the resulting causal
set in the above sense does not immediately follow. Consider by analogy a game of fortune
in which a circular wheel is spun to a random orientation. While the distribution of final
directions is indeed rotationally invariant, a particular outcome of the process is certainly
not. (A form of ``spontaneous symmetry breaking'', perhaps.) Likewise, a particular
outcome of the Poisson process might be able to prefer a frame, even though the process
itself does not.

So, the question becomes: Is it possible to use a sprinkling of Minkowski space to
select a preferred frame? We will prove a theorem that answers ``no'' to this question.
In fact, it answers the slightly more general question whether a sprinkling can pick out
a preferred time-direction (which is certainly possible if an entire frame can be derived.)
Below, we formalise the notion of deriving a direction from a sprinkling, and we prove a
theorem showing that this cannot be done. In this sense, the situation with sprinklings
of Minkowski space is even more comfortable than that with sprinklings of Euclidean
space. It is possible to associate a direction from the rotation group to a point in such a
sprinkling, as discussed later (although this will not stop anyone from maintaining that a
gas behaves isotropically in the continuum approximation; these locally defined directions
have little significance at that level), but the non-compactness of the Lorentz group makes
the Lorentzian case different.

Based on the theorem, we can assert the following. Not only is the Poisson process in
Minkowski space Lorentz invariant, but the individual realizations of the process are also
Lorentz invariant in a definite and physically important sense.\cite{bhs}
\end{quotation}

For their theorem, BHS consider $n$-dimensional Minkowski space
$M^n$, with a fixed point $O$, the origin. 
They then consider the action of $L_0$, the connected component of
the identity in $O(n - 1, 1)$, on $M^n$ with fixed point $O$. 
They define $\Omega$ to be the set of possible sprinklings in $M^n$, 
denoting the sprinkling Poisson process by $( \Omega, \Sigma , \mu )$, 
where $\mu$ is the probability measure on $\Omega$ and $\Sigma$ 
is the $\sigma$-algebra of all measurable subsets of $\Omega$.
They state that the measure is invariant under $L_0$, i.e.
\begin{equation}
\mu = \mu \circ \Lambda \, , \qquad {\rm for~all~} \Lambda \in L_0
\, , 
\end{equation}
adding the gloss that 
\begin{quote}
the probability of a
(measurable) set of possible sprinklings is the same as that of the set obtained applying
a Lorentz transformation to it. \cite{bhs}
\end{quote}

They then consider hypothetical maps $D: \Omega \rightarrow H$ from
the set of sprinklings to the hyperboloid $H$ of unit time-like
vectors in $M^n$.    They argue that if such a map defines a preferred timelike
direction for each sprinkling in a way that genuinely depends
only on the sprinkling (and so does not required a preferred frame
or other data for its definition), then it must be equivariant under the Lorentz
group: 
\begin{equation}
 D \circ \Lambda = \Lambda \circ D \qquad {\rm for~all~} \Lambda
\in L_0 \, . 
\end{equation}
But if $D$ is measurable, then $\mu \circ D^{-1}$ defines a
probability measure on $H$ that is invariant under $L_0$. 
Since $H$ is non-compact and has infinite volume, they argue, 
no such measure can exist, and hence their theorem follows. 

\section{Sprinklings or sprinklable sets?} 

From here on we focus on the case of sprinklings in $4$-dimensional Minkowski space,
unless explicitly specified otherwise, i.e. we take the manifold
$M=M^4$.   Our discussion of sprinklable sets applies equally well to $M^n$ for any $n \geq
2$; our discussion of the lacuna in the BHS theorem applies equally
well to $M^n$ for any $n \geq 3$.    We will discuss the definition and properties of sprinklings with respect to
global coordinates defining an inertial frame, taking $c=1$.\footnote{Of course, one {\it
  could} still consider local coordinate choices in Minkowski space, but
this complicates the discussion and offers no obvious advantage.}

Fixing coordinates $(x,y,z,t)$ gives a useful way of visualizing the
sprinkling process.  One can imagine small boxes defined by coordinate
increments $dx, dy, dz, dt$, with $4$-volume $V = dx dy dz dt$.
Following DHS, we can take the probability of a point being ``sprinkled'' into each box
to be $V /V_{\rm fund}$, and define the sprinkling Poisson process
to be the limit of this process as $dx, dy, dz, dt$,  and hence $V$, 
tend to zero.  
However, DHS argue, because the $4$-volume is Poincar\'e invariant, the limiting
distribution is independent of the coordinate choice: 
it is defined by the property that the
probability of finding a sprinkled point in any small $4$-volume $V$ is 
approximately proportional to $V/V_{\rm fund}$ and that these events
are independent.   

It is nonetheless difficult to define a notation for individual sprinklings
without fixing coordinates.   
Given coordinates $\{x, y, z, t\}$, we can, for example describe a
sprinkling
$S$ as an ordered list, $S = \{P_1 , P_2 , \ldots , P_n , \ldots \}$, where 
$P_n = (x_n , y_n , z_n , t_n )$ is chosen so that 
$l_1 \leq l_2 \leq \ldots \leq l_n \leq \ldots$, where 
$l_n =  ( x_n^2 + y_n^2 +z_n^2 + t_n^2 )^{1/2}$, with some
tie-breaking condition if any of the $l_k$ are equal.
The ``Euclidean distances from the origin'' $l_n$ have
no fundamental geometric significance, but define a useful
labelling.  

This gives a concrete way of describing sprinklings related by Lorentz
transformations, as in the BHS theorem. 
The sprinklings $S = \{P_1 , P_2 , \ldots , P_n , \ldots \}$ and
$S' = \{P'_1 , P'_2 , \ldots , P'_n , \ldots \}$ 
are on the same Lorentz orbit, $S' = \Lambda S$, if and only
if there is a bijection $ \rho: {\Bbb Z}^+ \rightarrow {\Bbb Z}^+$
such that $P'_i = \Lambda P_{\rho (i) }$ for all $i \in {\Bbb Z}^+$. 
    
But it also raises a concern: does DHS sprinkling in fact give
a well-defined probability distribution on the class of countably discrete sets
that are equipped with a Lorentzian distance function and a causal
structure, and that could isometrically be embedded in $M^4$? 
Or does it give something subtly different: a probability 
distribution on discrete subsets of $M^4$ defined with respect to 
some set of coordinates, whose choice breaks Poincar\'e invariance?

It might be objected here 
that, even if labels for sprinkled points are necessary in order to
distinguish sprinklings, they need not be defined by a coordinate
system.   One could, for example use a bijection between 
${\Bbb R}$ and $M^4$ to label each point in Minkowski space
by a real number, in a highly discontinuous way.\footnote{I 
thank Rafael Sorkin for this suggestion.}
Even then, to exclude the possibility that Poincar\'e invariance
is effectively broken by the labelling, one would need to show that there is 
no natural definition of coordinates implied by such a bijection.
In any case, a question would still remain: does DHS sprinkling give
a probability 
distribution on discrete subsets of $M^4$ defined with respect to 
some labelling?

We can put these questions another way.   Suppose that $S \in \Omega$ is an
outcome of a sprinkling process in $M^4$, defined with respect to a given origin
and frame.   Let $L_0$ be the subgroup of the Lorentz group that
preserves the causal ordering, and let $P_0$ be the subgroup of the Poincar\'e group
generated by $L_0$ and space-time translations.   Suppose that $\Pi \in P_0$ is a Poincar\'e transformation with $S \neq \Pi
S = S'$.\footnote{It is not essential to the discussion at
this point, but note that we should expect this last condition to hold
almost surely for any given nontrivial $\Pi$, given  
a well-defined probability distribution on sprinklings that justifies
the intuition that they are locally random.   Any such distribution should
assign measure zero to the set of sprinklings that have 
a non-trivial Lorentz automorphism.}  
Then do we treat $S$ and $S'$ as two distinct
possible outcomes of the sprinkling process, because their points have
different coordinates or labellings?   Or do we treat them as identical, representing a single outcome, because they are 
isometric and have identical causal structures?    

We need a separate terminology for this second option.
Define a {\it sprinklable set} $\tilde{S}$ to be a countably infinite set of points $\{ P_i \}$
on which a distance function $d(P_i , P_j ) \in {\mathbb R}$ is defined
and a causal relation $\prec$ is also defined, such that there exists a {\it causal isometry} -- an isometry
that also preserves the causal relation -- between $\tilde{S}$ 
and some sprinkling $S$ in $M^4$.  

\subsection{Sprinklable sets and sprinklable causal sets} 

A sprinklable set thus also defines a causal set, by
retaining the causal ordering $\prec$ but ignoring the
quantitative distance function $d$. 
We could define a {\it sprinklable causal set} $C$ to be a countably
infinite set of points $\{ P_i \}$ on which a causal relation $\prec$ is
defined such that there exists a {\it causal map} -- a map that 
preserves the causal relation -- between $C$ and some sprinkling
$S$ in $M^4$.   This may be the more fundamentally relevant   
definition for standard causal set theory, which focusses on
the causal ordering between points and does not assume a 
distance function.   However, we are interested in general
approaches to discretizing Minkowski space, which might also
include a distance function between the discretized points.    
Since the distance function is defined on sprinklings, we 
will keep open the possibility that it might be physically
significant, and focus on sprinklable sets rather than
sprinklable causal sets.    Every sprinklable set defines
a unique sprinklable causal set, but not vice versa.\footnote{
For example, a dilation $D\tilde{S}$ of the sprinklable set
$\tilde{S}$, defined with distance function $D d(P_i , P_j )$
for some fixed $D>0$, corresponds to the same sprinklable
causal set as does $\tilde{S}$.   
That said, if the sprinkling comes from a Poisson
process, then the set of sprinklable
sets with any density other than the given sprinkling density is
measure zero. 
One can also find other
constructions of distinct sprinklable sets corresponding
to the same sprinklable causal set.  Consider for example
a sprinklable set of the
form $\tilde{S} = S^+ \cup S^-$, where $S^{\pm}$ are disjoint
sets with the property that $s_- \prec s_+ $ for every $s_- \in S^-$
and $s_+ \in S^+$.   We can add a uniform timelike distance $T$ between
$S^-$ and $S^+$, by defining $d' (s_- , s_+ ) = d (s_- , s_+ ) + T$,
for some constant $T>0$ and all $s_- \in S^-$
and $s_+ \in S^+$, and set  $d' (s_- , s'_- ) = d (s_- , s'_- )$ 
for all $s_- \in S^-$
and $s'_- \in S^-$
and  $d' (s_+ , s'_+ ) = d (s_+ , s'_+ )$ 
for all $s_+ \in S^+$
and $s'_+ \in S^+$. 
The resulting sprinklable set $\tilde{S'}$ corresponds to the
same sprinklable causal set as does $\tilde{S}$.   
Again, though, this construction involves sprinklable sets 
belonging to a measure zero set of sprinklings. 
It is interesting to ask whether, in some sense to be rigorized, 
a sprinklable causal set arising from a Poisson sprinkling process
defines a unique sprinklable set with probability one.
I thank Fay Dowker for raising this question.}

We note below that the natural probability measure  
on sprinklable sets assigns only probability values $0$ and $1$. 
We argue that this is problematic, since it makes the hypothesis
of discrete Poincar\'e invariance only very weakly testable.  
This issue would also arise if we took sprinklable causal sets to 
be the fundamental objects.  The properties of 
the natural $\sigma$-algebra and probability measure on sprinklable
causal sets would also need separate analysis.  
We thus focus on sprinklable sets from now on.

\subsection{Sprinklable sets, sprinklings and Poincar\'e invariance}

If there is a causal isometry $ \phi: \tilde{S} \rightarrow S$ 
from a sprinklable set $\tilde{S}$ to a sprinkling $S$, and $\Lambda \in L_0$, then 
$\Lambda \circ \phi : \tilde{S} \rightarrow S'$ defines
a causal isometry between $\tilde{S}$ and the sprinkling $S' = \Lambda S$. 
The definition of sprinklable set is 
thus independent of any frame choice for $M^4$.  
Similarly, $\tau \circ \phi: \tilde{S} \rightarrow  S''$ defines
a causal isometry between $\tilde{S}$ and the sprinkling $S'' = \tau
S$, where $\tau$ is a space-time translation.    

Treating sprinklings as the fundamental objects
seems to undercut the case for sprinkling as 
a Poincar\'e invariant construction of discrete sets that could
substitute for continuous space-time manifolds as a fundamental arena
in which physics takes place.   
If sprinklings have to be understood as sets of points embedded in $M^4$, 
then our fundamental description of physics still involves 
a continuous space-time.   If they also need to be 
defined with respect to fixed coordinates or a labelling, 
in order to distinguish causally isometric sprinklings, 
then our fundamental description of physics also involves
either a preferred frame or a labelling.   Even if these are undetectable
by experiment, they remain present in the definition
of the sample space $\Omega$.\footnote{As John Bell commented in
a related context, this would seem an eccentric way to make
a world.\cite{bell1986beables}}

Treating sprinklable sets as the fundamental objects
in a physical theory may thus look more promising.   
However, to define a theory based on randomly chosen sprinklable
sets we need to define a probability space of these sets
without referring to any preferred 
frame or origin.
This requires us to define a $\sigma$-algebra and
probability measure on the sample space of sprinklable sets,
without using the frame-dependent definitions for 
$\Sigma$ and $\mu$.  

A natural choice seems to be to 
construct definitions that can be inferred from, but
defined independently of, those of $\Sigma$ and $\mu$, 
by considering an sprinklable set as a
coset of the Poincar\'e group acting on $\Omega$.
We will sketch such definitions below. 
They give us $\sigma$-algebra elements of measure zero -- for example
singleton sets $\Sigma_S = \{ \tilde{S} \}$
containing individual sprinklable sets
$\tilde{S}$. 
They also give us $\sigma$-algebra elements of measure one --
for example the set $\Sigma_{x,y}$ of all sprinklable sets
containing at least one pair of points $P_i , P_j $
such that $ x < d(P_i , P_j ) < y$ for any given pair
of real numbers $x<y$.  
However, the $\sigma$-algebra contains no elements with measures 
between $0$ and $1$.   We argue below that this leads to 
problems in understanding the physical implications of
theories in which sprinklable sets are fundamental objects
and in particular in testing such theories.  
First, we reconsider the BHS theorem and note a lacuna. 

\section{A lacuna in the BHS theorem}

In Ref. \cite{bhs}, isometric sprinklings $S$ and $S' = \Lambda S$,
for nontrival $\Lambda \in L_0$, are treated as distinct. 
This means that an algorithm for associating timelike directions to sprinklings
should define an equivariant map $D: \Omega \rightarrow H$ from
the set of sprinklings to the hyperboloid $H$ of unit time-like
vectors in $M^4$, and it is argued that no measurable equivariant 
maps exist.  

If we take sprinklable sets, rather than sprinklings, to be the fundamental
physical objects, then this argument needs reconsidering.  
The $\sigma-$algebra and probability measure on sprinklable sets
allow us to infer that, with probability one, a sprinklable set is represented by sprinklings that
have no Lorentz (or Poincar\'e) automorphism. 
Consider a hypothetical algorithm that is defined on sprinklable sets with no
Poincar\'e automorphism and that with probability one produces a
preferred timelike direction.   This must be defined in terms of intrinsic
properties of the sprinklable set: the positive or negative distances
between its points and their causal relationships.  

(An example of an algorithm defined in terms of intrinsic properties
would be to choose the pair of points ($P_1 \prec P_2$) with smallest positive timelike
separation, take $P_1$ the origin, and take the vector $P_1 P_2$ to
define the preferred timelike direction.   However, the probability of
a sprinklable set having a pair of points with smallest positive
timelike separation is zero, so this algorithm is almost never
well defined.) 

Such an algorithm would also define an algorithm that 
associates a preferred timelike direction to
a generic sprinkling, since we can consider the sprinkling as 
a sprinklable set if we ignore its specific embedding in $M^4$.  
The BHS theorem shows that no such algorithm exists, and hence
that no probability one algorithm for associating preferred
timelike directions to sprinklable sets can exist. 
Any algorithm defined on a measurable set of sprinklable sets
thus can only be defined on a set of measure zero, since (as we
discuss below) the $\sigma$-algebra contains only sets of measure one
and zero.   

One might, though, query whether it is so reasonable to restrict
attention to algorithms defined on measurable sets of sprinklable
sets, precisely because the $\sigma$-algebra is so relatively
sparse.   In any case, there is another issue with the BHS theorem: it
proves less than required.
Showing that there is no algorithm 
defining a measurable map from sprinklings to timelike directions
is necessary but not sufficient to establish that a typical
sprinkling is effectively Lorentz invariant in all the physically
relevant senses.   

\subsection{Partial breaking of Lorentz invariance in sprinklings}

Recall again BHS's comparison of sprinklings in Euclidean and
Minkowski space.  As BHS note, {\it given a fixed point $P$}, there
is a mathematically well-defined construction of a preferred
direction from a Euclidean sprinkling, given by taking the nearest
sprinkling point to $P$.   It is important to be clear about the
logic here: given some data, which would not break rotational
invariance in the continuous manifold, there is a mathematical
sense in which rotational invariance is broken in the discrete
approximation.  The point of BHS's theorem is to show that this
does not happen in the Minkowski case: given a fixed point $P$,
BHS argue, there is still not a mathematical construction 
that breaks Lorentz invariance.  

But more is needed.   One needs to show that, given any data
that leave some continuous subgroup of the Lorentz group
as a symmetry in the continuous case, there is no mathematical
construction that breaks this symmetry in the discrete case.  
To see this is not the case, suppose we are given 
two timelike separated points, $P \prec Q$.
In the continuous case, this breaks translation invariance, and partially breaks Lorentz
invariance, but leaves invariance under the spatial rotation
subgroup. 
In the discrete case, however, it allows mathematical constructions
that break the spatial rotation
invariance. 
For example, suppose the two given timelike separated points, $P \prec
Q$, belong to a sprinkling $S$.   
Let the line $PQ$ define
the axis for a time coordinate.   Now identify the point
$X \in S$ that has time coordinate between those of
$P$ and $Q$ and attains the minimum spatial separation
from the line $PQ$ (measured at equal times) among all
points in $S$.   That is, if we denote the time coordinate
by $x_0$, then $x_0 (P ) < x_0 (X) < x_0 (Q)$, and if 
$X' (X) $ is the point on $PQ$ with $x_0 (X') = x_0 (X)$, 
then $X$ is chosen to minimize $d(X,X'(X))$ over all
$X \in S$. 
This defines $X \in S$ uniquely except for a measure zero
subset of the sprinklings.\footnote{We could impose further
tie-breaking conditions to define $X$ on a still larger measure
one set.}
We can (for all but a measure zero subset of sprinklings) 
then associate the spatial direction $X'X$ 
to the combination $(S,P,Q)$ --- i.e. to the sprinkling
together with two given timelike  separated points. 

Another way of constructing a set of preferred directions 
from timelike separated points $P \prec Q$ is to consider the
longest chain $P\prec X_1 \prec \ldots \prec X_n \prec Q$ in the set,
assuming there is a nontrivial chain, and with some 
tiebreaking conditions\footnote{For example, we could choose
the chain for which $| d(P, X_1 )|$ is largest; if more than
one chains attain the maximum, then choose the chain amongst
these for which $| d(X_1 , X_2 ) |$ is largest, and so on.} 
if more than one chain attains
the maximal length.  This defines preferred timelike
vectors $PX_1 , X_1 X_2 , \ldots , X_n Q$, and 
a variety of constructions can be used to define preferred
spacelike vectors from these.   

\subsection{Are these constructions physically relevant?}

Whether these or other similar constructions might be physically
significant in a fundamental theory is an interesting question, whose
answer presumably depends on precisely which types of 
fundamental theory are considered.   For example, it seems a priori
conceivable that a dynamical theory on sprinklings
might imply that particles propagating from $P$ to $Q$ 
would cause observable anisotropic effects associated with
preferred directions selected by one of the rules above,
thus giving empirical evidence of the violation of local
Lorentz invariance.  (Recall that DHS's criterion 
requires that the underlying discreteness in a physical theory whose
approximating continuum is a Lorentzian manifold $M$ should
not suffice to distinguish even a {\it local} Lorentz frame at any point
of $M$.)  On the other hand, it might also be possible to
characterise interesting classes of theory for which BHS's 
analogy with Euclidean sprinkling models of a gas holds good,
in that such locally defined directions exist but can be
shown to have no global significance.  
Such a result would be very interesting, albeit weaker 
than BHS and DHS's claim that Lorentz invariance in 
{\it all} physically meaningful senses follows from the 
sprinkling construction.    

Whatever the physical status of the constructions, we believe the logical point
is clear.  The BHS theorem is intended to give a mathematical proof that Lorentz invariance
cannot be broken, as distinct from a physical argument that 
Lorentz invariance breaking is implausible.    
As BHS note, it is not possible to prove a version of their
theorem in Euclidean space, because a generic point in a Euclidean
sprinkling has a nearest neighbour, and so a choice of origin
generically allows a preferred direction to be defined.   
Although the constructions above are not as simple, they 
break Lorentz invariance
in a roughly analogous way, and so block the path to
rigorously proving full Lorentz invariance by any argument like that
of BHS.   While the BHS theorem applies to sprinklings, the same
argument applies for sprinklable sets, since our constructions
use only intrinsic properties.   

\section{Poisson processes on the real line}

To illustrate the properties of sprinklings, sprinklable sets and
their probability distributions, it is helpful to consider
a simpler example than sprinklings in Minkowski space. 

\subsection{Poisson processes on the real numbers}

We first consider the real numbers  $\mathbb R$ with their standard structure:
a preferred origin ($0$), a preferred direction (positive) and a
metric ($| x-y|$) that together allow us to define a signed distance function ($d(x,y) = y-x $).   This is the analogue of $M^4$ with
preferred coordinates ($x,y,z,t$), a preferred orientation
(distinguishing past and future timelike vectors), and the 
pseudo-Riemannian metric.  (Note though that, while the
pseudo-Riemannian metric does define positive and negative distances,
it is the analogue of the metric $| x-y|$, not of the signed
distance function $d(x,y)$.  The latter has no good analogue in
Minkowski space.) 
 
We define sprinkling on  $\mathbb R$ as  a random generalised Poisson process that selects a countable
set $S$ of points, with expected separation $D$ between
neighbouring points.   One way to define the probability distribution
on sprinklings $S$ is as follows.
First, we take a sprinkling to include all the points $x_1 < x_2 < \ldots$ generated by
a Poisson process with mean $D$ defined on $\mathbb R^+$. 
Then we define a second Poisson process with mean $D$ starting
at the point $x_1$ and extending in the negative direction, 
giving us points $\ldots x_{-1} < x_0 < x_1$.  
Our sample space of sprinklings is then
$\Omega = \{ \{ x_i \}_{i \in \mathbb Z} \}$, the set of 
ordered subsets of $\mathbb R$ labelled by the integers.   

For $n \in {\mathbb N}$, let $F_n$ be the
set of subsets of $\Omega$ of the form
\begin{equation}
\{ \, x_i \in \left[ a_i , b_i \right] \, : \, -n \leq i \leq n \, \} \, , 
\end{equation}
where $ a_i < b_i$ for each $i$.  
Let $F_{\infty} = \cup_{i \in {\mathbb N}} F_i$ be the collection of
subsets of $\Omega$ that can be defined by some finite
number of statements about the location of points $x_i$ in
finite intervals, and let $F= \sigma (F_{\infty} )$ be the
$\sigma$-algebra generated by $F_{\infty}$.  
Define the function $\mu_0$ on $F_{\infty}$ 
to be the probability that the generalized Poisson process
assigns the relevant points to the relevant intervals. 
This extends to a probability measure $\mu$ on $(\Omega , F_{\infty}
)$.

We can treat this as a toy model
of a physical universe, giving us a toy theory $T_0$ that makes non-trivial
probabilistic predictions.    For example, given any point of 
a sprinkling $S$ with coordinate $x$, the probability density for the next point 
in the positive direction having coordinate $x + y$ is 
$(\frac{1}{D})\exp ( - y/D )$.   

Given finite sets of data about sprinkling points, we can test theory $T_0$ against others.
For example, consider the deterministic
theory $T_1$ that predicts that points will be found precisely
at the locations $D {\mathbb Z}$, and the 
theory $T_2$ which predicts that the separations between points
are uniformly distributed on the interval $\left[ 0 , 2D \right]$. 
Suppose that we are somehow presented with a window onto the 
toy universe which exposes the first six points with
positive coordinates, $\{ x_1 , \ldots , x_6 \}$, and 
gives us these coordinates to infinite precision. 
The theory $T_1$ is excluded unless $x_i = D i$ for $i=1, \ldots , 6$. 
On the other hand, if this condition does hold, $T_1$ is 
effectively confirmed compared to the other theories,
which assign probability zero to this precise configuration. 

The relative probabilities of $T_0$ and $T_2$ are 
given respectively by
\begin{equation}
(\frac{1}{D} )^6  \exp ( - x_1 / D) \prod_{i=2}^6 \exp - ( (x_i - x_{i-1} ) /  D) \,
\end{equation}
and
\begin{equation}
( \frac{1}{2D} )^6 \theta ( 2 D - x_1 ) \prod_{i=2}^6 \theta ( 2 D -
x_i + x_{i-1} ) \, .
\end{equation}
If the second of these is zero, $T_2$ is also excluded.
Otherwise, if we had non-dogmatic Bayesian priors for 
$T_0$ and $T_2$ before seeing the data, these are rescaled
by the relative probabilities, but remain non-dogmatic.  

We could, of course, carry out similar calculations if we
were given intervals for the $x_i$ rather than precise
coordinates.  For example, if $x_i \in \left[ Di - \epsilon, Di + \epsilon \right]$
for $i = 1, \ldots , 6$, where $ \epsilon \ll D$, then 
$T_1$ is strongly favoured compared to $T_0$ and $T_2$, 
but neither of these are completely excluded.   

So, if we, as local observers in this one-dimensional discrete
toy universe, obtain evidence about the discretization in our local
neighbourhood that makes the Poisson sprinkling hypothesis statistically 
improbable compared to others, we have prima facie
justification for disfavouring it.   As in any cosmological model, the inference could possibly
be complicated by anthropic reasoning: if we had good 
reason to believe that observers tend to be located only in 
atypical regions of a Poisson sprinkling then we might readjust 
our inferred weights.   But scientific inference works as well
as one can hope in this type of toy model. 

\subsection{Poisson processes on a line with no fixed origin}

Now consider a translation invariant version of the previous Poisson 
sprinkling model in which $\{ x_i \}$ and $\{ x_i + x \}$ 
(for any real number $x$) are identified as the same outcome.
In other words, we have an action of the translation group ${\mathbb
  R}$ given by $t_x : \{ x_i \} \rightarrow \{ x_i + x \}$, and 
all the sprinklings in each coset are identified as a single outcome. 
Our toy universe is now described by some countable ordered sequence of points on 
a one-dimensional Riemannian manifold which is isometrically
isomorphic to ${\mathbb R}$, and has a preferred positive direction,
but which has no fixed reference point.    
The sample space $\Omega'$ consists of countably infinite unlabelled ordered sequences of 
points $\underline{P}$, between which a relative separation $d(P , Q ) \in
{\mathbb R}$ is defined, reflecting the isometry with ${\mathbb R}$
and the preferred positive direction.
Thus we have $d(P, Q) = - d(Q , P)$ and $d(P , R ) = d(P , Q ) + d ( Q , R )$.

We now need a $\sigma$-algebra and probability
measure defined on the cosets of Poisson sprinklings under
the translation group. 
One approach is to define sets that have specified properties
with respect to some (arbitrarily) chosen point $P$ in the set. 
To define a notation to describe these properties we label the point $P=P_0$, 
and label other points by their proximity to $P$; thus $P_{\pm 1}$
are the closest points to $P$ in the positive and negative directions,
and so on.    

For example, define $F'_n$ to be the set of 
subsets of $\Omega'$ that can be defined by the property that 
they contain a point $P = P_0$ whose nearest $n$ neighbouring 
points on either side, $P_{-n} < \ldots < P_0 \ldots
< P_n$, have separations lying in specified finite intervals.  
Thus an element of $F'_n$ takes the form
\begin{equation}
F ( a_{-n}, b_{-n} ; \ldots ; a_{n-1}, b_{n-1} ) = \{ \, \underline{P}
\, : \exists P = P_0 {\rm ~such~that~} \, d(P_{-n} , P_{-n+1} ) \in \left[ a_{-n} , b_{-n} \right]
\, , \ldots , d(P_{n-1} , P_n ) \in \left[ a_{n-1} , b_{n-1} \right] \, \} \, , 
\end{equation}
where $0 \leq a_{i} < b_{i}$ for each $i$. 
Define 
\begin{equation}
F'_{\infty} = \bigcup_{i \in {\mathbb N}} F'_i \, , 
\end{equation}
and let $F'= \sigma (F'_{\infty} )$ be the
$\sigma$-algebra generated by $F'_{\infty}$.  
Now the probability measure $\mu'$ on $( \Omega', F' )$ 
has 
\begin{equation}
\mu' ( F ( a_{-n}, b_{-n} ; \ldots ; a_{n-1}, b_{n-1} )) = 1 \, , 
\end{equation}
since almost every sequence
of points contains some finite subsequence of neighbouring points
whose separations lie in any given sets of finite intervals.   

We may enlarge the definition of $F'$, so as 
to include sets with well-defined asymptotic properties.   
For example, we may include sets  
\begin{equation}
B_p = \{ \, \omega \in \Omega' \, : \, 
\exists P = P_0 {\rm~such~that~}\lim_{ n \rightarrow \infty } \frac{ d(P_{-n} , P_n ) }{ 2n }
= \delta  \, \} \, .
\end{equation}
If this asymptotic property holds for one point $P=P_0$ in the set
then it holds for every point, so we could also write 
\begin{equation}
B_p = \{ \, \omega \in \Omega' \, : \, 
\forall P \in \omega {\rm~if~}P=P_0 {\rm~then~} \lim_{ n \rightarrow \infty } \frac{ d(P_{-n} , P_n ) }{ 2n }
= \delta \, \} \, . 
\end{equation}
The probability measure has $\mu' (B_p ) = 0$ for $\delta \neq D$ and
$\mu' (B_p ) = 1$ for $\delta = D$. 

We can also include sets with more complicated limiting properties,
so that for example we can justify the statement that 
\begin{quote}
If a sequence of points $\underline{P}$ is partitioned into length $(2n+1)$ subsequences 
in any of the $(2n+1)$ possible ways, then, with probability one,
the asymptotic proportion of subsequences $X_1 \ldots X_{2n+1}$
with separations \mbox{$d(X_1, X_2 ) \in \left[a_1 , b_1 \right ] , \ldots ,
d(X_{2n} , X_{2n+1} ) \in \left[a_{2n} , b_{2n} \right]$}
is 
\begin{equation}\label{sepprob}
\prod_{i=1}^{2n}  ( \exp (-a_i / D) - \exp (-b_i / D ) ) \, .
\end{equation}
\end{quote} 

These sets all have probability measure $0$ or $1$; it follows that
their countable unions and intersections also have measure $0$ or $1$.
Since no generating element of the translation invariant $\sigma$-algebra has non-trivial
probability, it seems intuitively clear that the translation invariant probability
measure assigns only trivial probabilities to 
all elements; Ref. \cite{last2017lectures} gives a formal proof.\footnote{I
thank Rafael Sorkin for locating this reference.}

Suppose we are now somehow presented with a window onto 
this translation invariant toy universe,
which exposes a length $(2n+1)$
subsequence of neighbouring points to us, without assigning any coordinate labels
to the points.  
One would like to be able to say that there is probability 
$\prod_{i=1}^{2n}  ( \exp (-a_i / D) - \exp (-b_i / D ) ) $
that its separations obey $d(X_1, X_2 ) \in \left[a_1 , b_1 \right ] , \ldots ,
d(X_{2n} , X_{2n+1} ) \in \left[a_{2n} , b_{2n} \right]$. 
However, we now have an explanatory gap.   The translation-invariant
$\sigma$-algebra indeed contains a set of sequences defined by the property that
they contain such a subsequence, but it has probability one. 
More generally, since the probability measure assigns only trivial probability values 
to sets in the $\sigma$-algebra, we cannot infer any non-trivial
probability value about any proposition.  

In comparing toy model theories, all we can thus do is exclude
theories that assign probability zero to observed data.
Consider a translation invariant version of our 
previous example, in which our window gives us seven neighbouring points 
$P_0 , P_1 , \ldots , P_6$, with $d(P_{0}, P_i )  \in \left[ Di -
\epsilon, Di + \epsilon \right]$ for $i = 1, \ldots , 6$, where $ \epsilon
\ll D$.
Each of the theories $T_0$, $T_1$ and $T_2$, in their translation
invariant form, predicts that such sequences will arise with
probability one.   None is favoured over the others by 
our observation.  

\section{Are relative frequencies relevant?}

As we noted, although the translation invariant probability measure
does not assign non-trivial probabilities to any event, it does
assign probability one (or zero) to events defined by relative
frequencies.  If a sequence of points $\underline{P}$ is partitioned into length $(2n+1)$ subsequences 
in any of the $(2n+1)$ possible ways, then, with probability one,
the asymptotic proportion of subsequences $X_1 \ldots X_{2n+1}$
with separations \mbox{$d(X_1, X_2 ) \in \left[a_1 , b_1 \right ] , \ldots ,
d(X_{2n} , X_{2n+1} ) \in \left[a_{2n} , b_{2n} \right]$}
is given by Eqn. (\ref{sepprob}).  

There are certainly suggestions in the physics literature
that showing with probability one that an event has relative frequency
$p$ in an infinite sequence implies that the individual events
have probability $p$.  Or, at least, that careful analysis of the 
statement ``individual events have probability $p$'' leads to the 
conclusion that it means no more than that the relative frequency is
(or would be) $p$ in an infinite sequence. 
Discussions of many-worlds theories
includes arguments along these lines by Hartle \cite{hartle1968quantum},
Coleman \cite{coleman} and
Aguirre-Tegmark \cite{aguirre2011born}, among others.    

These suggestions tend to be aligned with frequentist views of
probability, which some find persuasive but which also have well known
problems: a good summary of arguments and critiques can be found in
Ref. \cite{sep-probability-interpret}.   
Among the points we think worth highlighting are that in general relative frequencies do not respect countable additivity;
moreover, the sets on which they are defined are not closed under countable
union, nor under finite intersection \cite{reichenbach1971theory, fine2014theories, kac1959statistical,
  suppes2002representation, van1979relative}.   
Relative frequencies of Poisson processes also illustrate a version of
Reichenbach's machine-gun example and its challenge to frequentists. \cite{reichenbach1971theory,
  van1979relative}.  To see this, consider an instance of a Poisson process
on a line with no fixed origin, and then consider the set $US$ of unrealised
separations between neighbouring points in this instance, i.e.~the positive real line minus the countable set 
 $ \{ \ldots, d(X_1 , X_2 ) , d(X_2 , X_3) , \ldots \}$. 
By definition, the event that a separation between neighbouring points
lies in $US$ has relative frequency zero in the realised instance: none 
of the separations lies in $US$. 
However, the a priori probability that a separation will lie in $US$, according to the
model, is one, since almost all real numbers belong to $US$.
This and the other issues highlighted apply equally well, of course, to 
the probabilistic model of sprinklable sets described above.   

Whatever view one takes on the problems of frequentism and probability in general, 
the fundamental issue here is simple.   If we define the relevant 
Poisson processes for sprinklable sets in the standard way, in terms of a sample space,
$\sigma$-algebra and measure, then we cannot derive non-trivial
probabilistic statements, since the probability measure only allows
trivial values.    Some frequentists (among others) might take
the view that {\it all} well-defined probabilistic statements
ultimately refer to probabilities zero or one: that statements
with intermediate probability values mean precisely that a 
relevant relative frequency takes a value with probability one.  
Even if defensible, this stance is not very helpful if our aim is to justify testing and
relative confirmation of a sprinklable set model on the basis of finite data.   If discrete space-times should
properly be modelled by sprinklable sets, but theories of
this type cannot be tested by any finite set of observations,
it is cold comfort that models of discrete space-times
as sprinklings cannot be finitely tested either.   

\section{Summary: probability and sprinklable sets} 

Does there exist a well-defined probabilistic theory in which the
fundamental physical objects are sprinklable sets
with a particular dimensionality (for example, $3+1$)?
An affirmative answer requires a rigorous definition
of the probability space of sprinklable sets.
So far as we are aware, no such definition has yet appeared
in the causal set literature.   However, the $\sigma$-algebra
and measure on Poincar\'e orbits of sprinklings appear 
natural candidates. 

A Poincar\'e invariant sigma algebra is a necessary precondition
for a model of discrete space-times that might reasonably be
said to respect Poincar\'e invariance, but it is not sufficient. 
To take an extreme example: the sigma-algebra  
\begin{equation}
G = \{ \emptyset , \Omega \} \, , 
\end{equation}
where $\Omega$ is the set of all countable subsets of $M^4$, has
a Poincar\'e invariant definition: both $\emptyset$ and $\Omega$
are Poincar\'e invariant sets.   The measure 
$\mu$ defined by $\mu ( \emptyset ) = 0$ and $\mu ( \Omega ) = 1$
gives us a well-defined probability space.   But no one should
claim that this defines a satisfactory physical model or
that it gives good reason to be optimistic about the
existence of discrete Poincar\'e invariance.    
The sigma algebra needs to have enough structure to 
allow us to derive whatever consequences are supposed to follow
from the theory.    These certainly should include the lack
of an observable preferred frame.  

One concern here is that, once 
we move from treating sprinklings as fundamental physical
objects to considering sprinklable sets as fundamental,
we lose the possibility of {\it proving} typical large-scale
properties of the sample set from the properties of local
probabilistic processes. 
Instead, we need to {\it postulate} the large-scale properties, 
and all of our postulates involve sets of measure zero or one.  
Essentially, we postulate that a large-scale property will
almost always or almost never hold, by choosing to include
in the sigma-algebra either the set of sprinkable sets that satisfy the property, or its
complement.    
So long as we respect the closure axioms for the sigma algebra, we
thus seem free to choose whether 
or not to adopt physically relevant postulates, such as those
describing the asymptotic density of the sprinklable set.   
It might be argued that 
this is more of an aesthetic concern
than a logical one, since all physical theories are based on some
postulates. Still, we should at least be clear
whether a proposed discrete theory has particular properties 
simply because we choose to impose them by fiat, or whether 
they are provable consequences of some simpler underlying structure. 

A stronger concern also arises from the fact that the natural probability measure
on sprinklable sets takes only values one and zero. 
It seems to follow from this that our credence 
in a theory can be altered only if observations produce data
that the theory assigns probability zero, in which case our
credence also becomes zero.   If so, no apparent
breakdown of Poincar\'e invariance (however dramatic) in any finite
region (however large) of
space-time gives any evidence against the hypothesis
that the full sprinklable set derives from a 
Poincar\'e invariant model.    
It seems that proponents of a fundamentally indeterministic theory of sprinklable sets
either have to accept that the hypothesis of discrete
Poincar\'e invariance is effectively untestable, or perhaps try to justify 
a role for non-trivial probabilities in 
a theory of the matter distribution on sprinklable sets.   

One possible fallback position is that the ultimate aim
is to find a discrete cosmological theory of matter and gravity, 
which may include physically preferred space-time points (such as
points modelling the initial singularity) or frames (such as the
cosmological centre-of-mass frame).   Whether any such theory
can retain local Lorentz or Poincar\'e invariance in a physically
meaningful and fundamentally significant sense is not obvious,
though. 

We also noted a lacuna in the BHS argument, which means that there is
as yet no 
rigorous proof that the typical sprinkling or sprinklable set is fully
Lorentz invariant.   As BHS note, their argument also fails for
sprinklings in Euclidean space, and this is not generally seen as
a disaster for claims that physical systems modelled by Euclidean
sprinklings are effectively isotropic.  Still, it removes some
of the comfort that a rigorous theorem would provide. 

Some issues formally similar to those considered
here arise in discussions
of many-worlds theories.   In particular, Aguirre--Tegmark's
``cosmological interpretation of quantum mechanics''
\cite{aguirre2011born}
relies on propositions about relative frequencies in an infinite collection
of unindexed universes.   These raise questions about the definition
of the probability space and the jump from propositions about
relative frequencies to statements about probabilities of individual
events, which appear very similar to those considered here. 
However, 
since Aguirre--Tegmark's cosmological models
are not perfectly
analogous to sprinklable set models, and since both cosmological
and Everettian many-worlds theories raise other issues that
need separate discussion, we leave this for future work.   

\vskip 10pt

{\bf Acknowledgments} \qquad 
I am very grateful to Fay Dowker and Rafael Sorkin for very helpful
discussions and explanations.  
This work was partially supported by an FQXi grant and by
Perimeter Institute for Theoretical Physics. Research at Perimeter
Institute is supported by the Government of Canada through Industry
Canada and by the Province of Ontario through the Ministry of Research
and Innovation.   
	
\bibliographystyle{unsrtnat}
\bibliography{sprinklings}{}

\begin{thebibliography}{15}
\providecommand{\natexlab}[1]{#1}
\providecommand{\url}[1]{\texttt{#1}}
\expandafter\ifx\csname urlstyle\endcsname\relax
  \providecommand{\doi}[1]{doi: #1}\else
  \providecommand{\doi}{doi: \begingroup \urlstyle{rm}\Url}\fi

\bibitem[Dowker et~al.(2004)Dowker, Henson, and Sorkin]{dhs}
Fay Dowker, Joe Henson, and Rafael~D Sorkin.
\newblock Quantum gravity phenomenology, {L}orentz invariance and discreteness.
\newblock \emph{Modern Physics Letters A}, 19\penalty0 (24):\penalty0
  1829--1840, 2004.

\bibitem[Bombelli et~al.(2009)Bombelli, Henson, and Sorkin]{bhs}
Luca Bombelli, Joe Henson, and Rafael~D Sorkin.
\newblock Discreteness without symmetry breaking: a theorem.
\newblock \emph{Modern Physics Letters A}, 24\penalty0 (32):\penalty0
  2579--2587, 2009.

\bibitem[Bedingham(2018)]{bedingham2018collapse}
Daniel~J Bedingham.
\newblock Collapse models and space--time symmetries.
\newblock In Shan Gao, editor, \emph{Collapse of the Wave Function: Models,
  Ontology, Origin, and Implications}, pages 74--94. Cambridge University
  Press, 2018.

\bibitem[Hossenfelder(2015)]{0264-9381-32-20-207001}
Sabine Hossenfelder.
\newblock A no-go theorem for {P}oincar\'e-invariant networks.
\newblock \emph{Classical and Quantum Gravity}, 32\penalty0 (20):\penalty0
  207001, 2015.
\newblock URL \url{http://stacks.iop.org/0264-9381/32/i=20/a=207001}.

\bibitem[Last and Penrose(2017)]{last2017lectures}
G{\"u}nter Last and Mathew Penrose.
\newblock \emph{Lectures on the {P}oisson process}, volume~7.
\newblock Cambridge University Press, 2017.

\bibitem[Hartle(1968)]{hartle1968quantum}
James~B Hartle.
\newblock Quantum mechanics of individual systems.
\newblock \emph{American Journal of Physics}, 36\penalty0 (8):\penalty0
  704--712, 1968.

\bibitem[Coleman(1994)]{coleman}
Sidney Coleman.
\newblock Quantum mechanics in your face. {L}ecture given at the {N}ew
  {E}ngland sectional meeting of the {A}merican {P}hysical {S}ociety, 1994.
\newblock URL
  \url{http://media.physics.harvard.edu/video/index.php?id=SidneyColeman_QMIYF.flv}.

\bibitem[Aguirre and Tegmark(2011)]{aguirre2011born}
Anthony Aguirre and Max Tegmark.
\newblock Born in an infinite universe: a cosmological interpretation of
  quantum mechanics.
\newblock \emph{Physical Review D}, 84\penalty0 (10):\penalty0 105002, 2011.

\bibitem[H\'ajek(2012)]{sep-probability-interpret}
Alan H\'ajek.
\newblock Interpretations of probability.
\newblock In Edward~N. Zalta, editor, \emph{The Stanford Encyclopedia of
  Philosophy}. Metaphysics Research Lab, Stanford University, winter 2012
  edition, 2012.

\bibitem[Reichenbach(1971)]{reichenbach1971theory}
Hans Reichenbach.
\newblock \emph{The theory of probability}.
\newblock Univ of California Press, 1971.

\bibitem[Fine(2014)]{fine2014theories}
Terrence~L Fine.
\newblock \emph{Theories of probability: An examination of foundations}.
\newblock Academic Press, 2014.

\bibitem[Kac(1959)]{kac1959statistical}
Mark Kac.
\newblock \emph{Statistical independence in probability, analysis and number
  theory}, volume 134.
\newblock Mathematical Association of America Oberlin, OH, 1959.

\bibitem[Suppes(2002)]{suppes2002representation}
Patrick Suppes.
\newblock \emph{Representation and invariance of scientific structures}.
\newblock CSLI publications Stanford, 2002.

\bibitem[Van~Fraassen(1979)]{van1979relative}
Bas~C Van~Fraassen.
\newblock Relative frequencies.
\newblock In M.H. Salmon, editor, \emph{Hans Reichenbach: logical empiricist},
  pages 129--167. Springer, 1979.

\bibitem[Bell(1986)]{bell1986beables}
John~S Bell.
\newblock Beables for quantum field theory.
\newblock \emph{Phys. Rep}, 137:\penalty0 49--54, 1986.

\end{thebibliography}

\end{document}